\documentclass[useAMS,usenatbib]{mn2e}
\usepackage{graphicx}
\usepackage{epstopdf}

\title[Time to first detection of double neutron star mergers ]{Observation time to first detection of double neutron star mergers by gravitational wave observatories }
\author[D. M. Coward]{D. M. Coward$^{1}$\thanks{E-mail:
coward@physics.uwa.edu.au} \\
$^{1}$School of Physics, University of Western Australia, M013, Crawley WA 6009, Australia\\}
\begin{document}

\date{Accepted Received ; in original form }

\pagerange{\pageref{firstpage}--\pageref{lastpage}} \pubyear{2008}

\maketitle

\label{firstpage}

\begin{abstract}
We constrain the uncertainty in waiting times for detecting the first double-neutron-star (DNS) mergers by gravitational wave observatories. By accounting for the Poisson fluctuations in the rate density of DNS mergers and galaxy space density inhomogeneity in the local Universe, we define a detection `zone' as a region in a parameter space constrained by the double neutron star merger rate and two LIGO operations parameters: an observation horizon distance and science run duration. Assuming a mean rate of about 80 DNS mergers per Milky Way galaxy Myr$^{-1}$, we find a 1/20 chance of observing a merger by Enhanced LIGO in only 1 yr of observation. The minimum waiting time and temporal zone width for an Advanced LIGO sensitivity are much shorter and imply that there is a 95\% probability of detecting a DNS merger in less than 60 days and a 1/20 chance of a first detection in about 1 day. At the 5\% probability threshold for a first detection, we find that the effect of galaxy clusters on detection is smoothed out and may only influence detection rates after 5--10 years observation time.

\end{abstract}


\section{Introduction}
Double neutron star (DNS) binary mergers are potentially 
strong sources of detectable gravitational wave (GW) emission. Nine DNS
binary systems have been discovered, eight of them are in the Galactic disk and one is in a globular cluster (Stairs 2004, Lorimer et al. 2006). Energy loss from GW emission \citep{Tayl89} causes an orbital
in-spiral until the binary system merges, resulting in a burst of GWs usually described as a `chirp' signal. 

The US LIGO (Laser Interferometer Gravitational-wave Observatory) is searching for, among other potential sources, DNS mergers in the local Universe and has extended its sensitivity horizon to these events to some tens of Mpc (Abbott et al. 2007). LIGO sensitivity to a DNS merger depends on its distance and relative orientation to the event. To aid searches a catalogue has recently been compiled of  distances, sky positions and blue luminosities of galaxies in the local Universe: the Compact Binary Coalescence Galaxy (CBCG) catalogue \citep{kopp08}.  

Theoretical calculations and simulations predict a wide range of delay
times between DNS formation and merger (see Belczynski et al. 2006). Their results 
support the argument that a significant fraction of DNS mergers follow massive star formation.
Blue luminosity is a known indicator of massive star formation and may also provide a means of tracking the DNS merger rate, for binaries with merger times that are short compared to the Hubble time. The relatively frequent occurrence of such systems is supported by the discovery of the double pulsar system J0737--3039A,B. With an orbital period of only 2.45 hr, it will coalesce in only 87 Myr. In the local Universe, out to some hundreds of Mpc, we assume, like most other studies, e.g. \cite{Phin91} and others, that the DNS merger rate scales with blue luminosity.

Interestingly, some short gamma ray bursts (GRBs), observed by the Swift satellite, may also be the high-energy EM emissions from DNS mergers. This is evidenced by the optical localization of several short hard bursts with their host galaxies. The identification of short GRBs with different types of galaxies and distances from galaxy centers is related, in a non-trivial way, to the DNS velocity and the delay times between formation and merger (see e.g. Belczynski et al. 2006). In contrast to the binary pulsar systems, the offset position for GRB050509b at 40 $\pm 13$ kpc from its elliptical galaxy G17 implies large kick velocities \citep{grind06}. This discrepancy between the observed radio pulsar binaries and short GRB locations could indicate that DNS mergers may occur in both old stellar populations, such as globular clusters, and relatively younger environments.  Even though the exact  distribution of DNS mergers with galaxy types is not well constrained, it is reasonable to assume that there is a strong correlation between massive star formation rates, DNS merger rate and the blue luminosity of the progenitors.

With the assumption that galactic blue luminosity is a direct measure of massive star formation in the local Universe, the CBCG catalogue can be used to infer the DNS merger rate at extra-galactic distances by scaling from the blue luminosity of our Galaxy. This requires knowledge of the Galactic DNS merger rate. In order to estimate the DNS coalescence rate, Kalogera et al. (2001) used a semi-empirical approach, based on the observed properties of known DNS systems and pulsar survey selection effects, to obtain scale factors that correct for the unobserved fraction of existing systems. With this model and more recent pulsar survey results that include J0737--3039, \cite{Kalog04} presented bounds for the merger rate for Galactic disk DNS systems, yielding $\mathcal R_{\mathrm{DNS}} = 83^{+209.1}_{-66.1}$ Myr$^{-1}$. \cite{Reg05}, using numerical simulations, find that mergers are possible between $2 \times 10^5$ yr and the age of the Universe. Using both evolutionary and statistical models, \cite{Reg05} find a Galactic merger rate of 17 Myr$^{-1}$,  similar to the lower bound calculated by \cite{Kalog04}. It is clear from these studies that the DNS merger rate is highly uncertain by several orders of magnitude. To account for this uncertainty, we employ a lower range of values for the DNS merger rate of $1-100$ Myr$^{-1}$ per Milky Way Galaxy.

\section{The model}
\subsection{DNS merger detection rates}
In a previous study, \cite{cow05} employed a simple Euclidean model for the rate density of DNS mergers to study the fluctuations in the detection rate of DNS by LIGO and Advanced LIGO. In this study, the aim is to show how the waiting time for a first detection of a DNS merger by next generation LIGO depends on the distribution of galaxies, the sensitivity of the detector and the observation time. 
To model the effect of the uncertainty in waiting times of DNS mergers, we start with a semi-empirical model to account for fluctuations in the cumulative merger rate arising from inhomogeneities in the local Universe. 

Following \cite{kopp08}, we employ the CBCG catalogue, using the cumulative sum of blue luminosities from the catalogue, $C_{L}(D)$, where $D$ is distance. We re-scale this galactic blue luminosity as a function of distance to detector horizon distance $D_{\mathrm{h}}$, \footnote{This distance is defined as the maximum distance at which LIGO can detect an optimally orientated DNS merger with signal-to-noise ratio of eight. For non-optimally orientated mergers, the sensitivity distance, $D_{\mathrm{s}}$, accounts for the antenna pattern of the detector, and is defined as $D_{\mathrm{s}}=D_{\mathrm{h}}/\sqrt{5}$}
so as to account for the response of the LIGO detector to GWs from a randomly oriented DNS merger averaged over all sky positions. The cumulative number of DNS mergers, $N_{\mathrm{DNS}}(D_{\mathrm{h}})$, in observation time $T$ detectable within horizon distance $D_{\mathrm{h}}<100$ Mpc is approximated using $C_{L}(D_{\mathrm{h}})$. For $D_{\mathrm{h}} > 100$ Mpc, the galaxy distribution is  considered uniform, so we extrapolate using a cubic power law, similar to eq. (9) in \cite{kopp08}. Thus, for $D_{\mathrm{h}}<100 \textrm{ Mpc}$,
\begin{equation}
N_{\mathrm{DNS}}(D_{\mathrm{h}}) = 
 1\times10^{-3} 
\Bigg(\frac{\mathcal R_{\mathrm{DNS}}}{L_{10}^{-1} \textrm{ Myr}^{-1}}\Bigg )
\Bigg(\frac{C_{L}(D_{\mathrm{h}})}{10^3 L_{10}}\Bigg)
\Bigg(\frac{T}{\mathrm{ yr}}\Bigg), 
\end{equation}
and, for $D_{\mathrm{h}} > 100 \textrm{ Mpc}$,
\begin{equation}
N_{\mathrm{DNS}}(D_{\mathrm{h}})\!
 = \! 4 \times 10^{-3} \!
\Bigg(\frac{\mathcal R_{\mathrm{DNS}}}{L_{10}^{-1} \textrm{ Myr}^{-1}} \Bigg)\!
\Bigg( \frac{D_{\mathrm{h}}}{100 \textrm{ Mpc}} \Bigg)^3 \!
\Bigg(\frac{T}{\mathrm{ yr}}\Bigg),
\label{nevents}
\end{equation}
where $L_{10}=10^{10} L_{B,\odot}$ is in units of solar blue luminosity.
\begin{figure}
\includegraphics[scale=0.75]{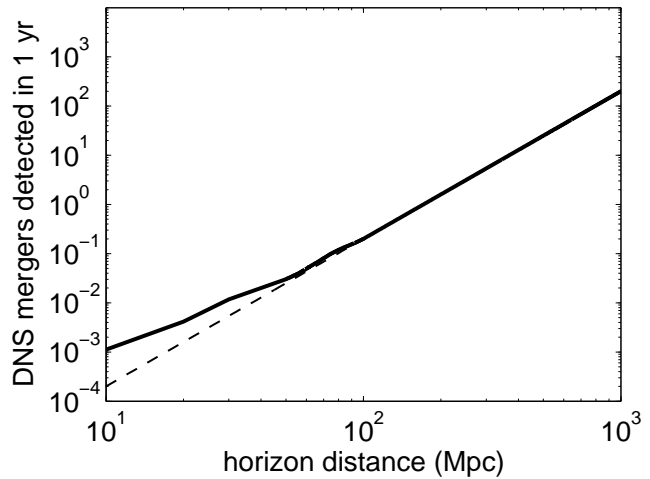}
\caption{Cumulative DNS mergers detected as a function of LIGO horizon distance assuming the most probable rate density of 83 mergers per Milky Way galaxy yr$^{-1}$ from Kalogera et al. (2004) and the scaling of blue luminosity to DNS number density from Kopparapu et al. (2008) who used the Compact Binary Coalesence Galaxy catalogue. We extrapolate the cumulative number  beyond 100 Mpc using a simple Euclidean cubic power law (dashed curve).} \label{fig1} 
\end{figure}

Figure \ref{fig1} plots equations 1 and 2 assuming an observation time of 1 yr and a Milky Way DNS merger rate of 83 Myr$^{-1}$, corresponding to the most probable rate from \cite{Kalog04}. For horizon distances comparable to that of Enhanced LIGO, $\approx60$ Mpc, the rate is about 1 per 20 years; for a horizon distance comparable to that of Advanced LIGO, $\approx450$ Mpc, the rate is about 20 per yr.

\subsection{Waiting times for detection of DNS mergers}
In addition to the uncertainty in the mean DNS merger rate, detection also depends on the intrinsic uncertainty in the waiting times between events. \cite{How07} demonstrated, using GRB peak flux data from the {\it Swift} satellite, that the waiting times of cosmic transients follow Poisson statistics. The study derived a power-law relation that is independent of the luminosity distribution of the sources to predict waiting times for bright GRBs observed by {\it Swift}. At the core of these models is the inclusion of the Poisson-based uncertainty in observation waiting times for astronomical transients. 

As in that work, we assume that DNS mergers are independent events, i.e. the occurrence of a DNS merger in one region of space does not influence the probability of another such event occurring. Therefore, the temporal separation between events detected by LIGO will follow an exponential distribution, defined by a mean event rate above $\mathcal R_{\mathrm{DNS}}(D_{\mathrm{h}})$ for events out to some horizon distance $D_{\mathrm{h}}$ in Euclidean space. The probability for at least one event  to occur in a volume bounded by $D_{\mathrm{h}}$ during an observation time $T$ at
constant probability $\epsilon$ is given by:
\begin{equation}\label{eq_peh}
\mathcal{P}(n \ge 1;\mathcal R_{\mathrm{DNS}}(D_{\mathrm{h}}),T)=  1 - e^{\mathcal R_{\mathrm{DNS}}(D_{\mathrm{h}})T} = \epsilon \,.
\end{equation}

\noindent For this equation to remain satisfied with increasing observation time:
\begin{equation}\label{eq_eps}
N_{\mathrm{DNS}}(D_{\mathrm{h}}) = \mathcal \mathcal R_{\mathrm{DNS}}(D_{\mathrm{h}})T =  |\mathrm{ln}(1 - \epsilon)|\;, \\
\end{equation}
with mean number of events $N_{\mathrm{DNS}}(D_{\mathrm{h}})$.

The probably event horizon (PEH) is defined by the minimum detectability horizon distance, $D_{\epsilon}^{\mathrm{PEH}}(T)$, for at least one event to occur over some observation time $T$, with probability above some selected threshold $\epsilon$. 
$D_{\epsilon}^{\mathrm{PEH}}(T)$ is defined by by fixing $\epsilon$ and solving the above relation, thus defining the PEH \citep{cow05}. 

In this study, we use two thresholds $\epsilon=(0.5,0.95)$, corresponding to 5\% and 95\% probability respectively of observing at least one event within $D_{\mathrm{h}}$. In other words we would be surprised to observe an event before $T(D_{5\%}^{\mathrm{PEH}})$, henceforth defined as the `waiting time', and extremely surprised if no events were observed by $T(D_{95\%}^{\mathrm{PEH}})$. For obvious reasons we denote the interval between these two times the `detection zone':
\begin{equation}\label{ezone}
 T_{\mathrm{zone}}(D_{\mathrm{h}})=T(D_{95\%}^{\mathrm{PEH}})-T(D_{5\%}^{\mathrm{PEH}})\;.
\end{equation}

\begin{table}
\caption{The waiting times for a first detection, in days, for the 5\% and 95\% probability thresholds assuming horizon and sensitivity distances, $D_{\mathrm{h}}$ and $D_{\mathrm{s}}$, corresponding to Advanced and Enhanced LIGO, along with the temporal zone widths.}
\begin{center}
\begin{tabular}{@{}lcccc}
\hline
 & $T_{\mathrm{5\%}}$ & $T_{\mathrm{95\%}}$ & $T_{\mathrm{zone}}$ \\
 \hline 
 Adv. LIGO $D_{\mathrm{h}}$ & 1 & 60 & 60 \\
 Adv. LIGO $D_{\mathrm{s}}$ & 10 & 600 & 590 \\
 \hline
 Enh. LIGO $D_{\mathrm{h}}$ & 365 & 21900 & 21800 \\
  Enh. LIGO $D_{\mathrm{s}}$ & 2000 & $<10^4$& $<10^4$ \\
\end{tabular}
\end{center}
\label{default}
\end{table}%

\begin{figure}
\includegraphics[scale=0.75]{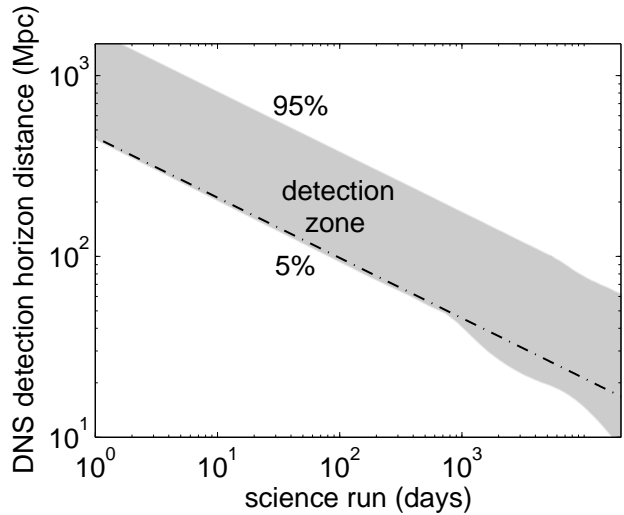}
\caption{The plot shows the evolution of the DNS merger detection zone (shaded region) bounded by the 5\% and 95\% probability contours for detection by LIGO. The intersection of a horizon distance (vertical axis) with the 5\% probability contour defines the waiting time before entering the detection zone. Assuming the most probable rate density of about 80 DNS mergers per Milky Way galaxy yr$^{-1}$ from Kalogera et al. (2004), the minimum length of the science runs to enter the zone is 365 and 2 days for Enhanced and Advanced LIGO sensitivities respectively. The detection zone width for Advanced LIGO is constrained to 2--60 days, while for Enhanced LIGO it is 1--60 yrs. The deviations of the probability contours from a simple power law, dot-dash curve, at horizon distances less than 50 Mpc is a result of fluctuations in the space density of galaxies and clustering in the local Universe (see Figure 3).}
\end{figure}\label{pehf}

\begin{figure}
\includegraphics[scale=0.75]{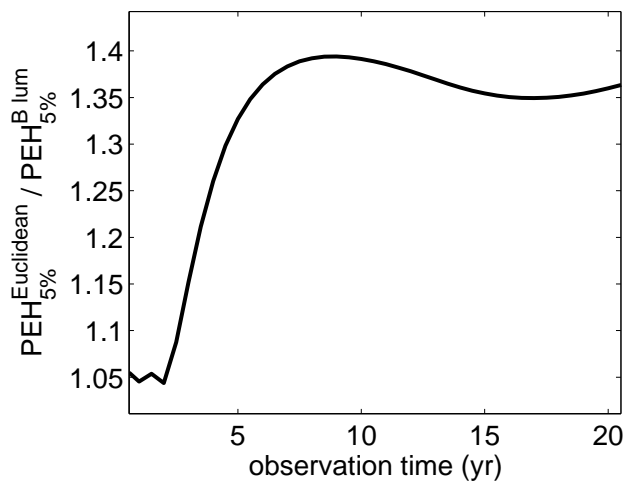}
\caption{Plot of the fractional difference between the 5\% detection probability contours (PEH$_{5\%}$) assuming a uniform distribution of galaxies compared to a model that includes fluctuations in the space density of galaxies in the local Universe. Clustering in the local Universe results in a maximum decrease of about 1.4 in the PEH$_{5\%}$ distance at an observation time of about 7--10 yr. For practical waiting times for a first detection i.e. 1-5 years, the effect from inhomogeneities of the galaxy distribution is not significant.}
\end{figure}\label{fig3}

\begin{figure}
\includegraphics[scale=0.75]{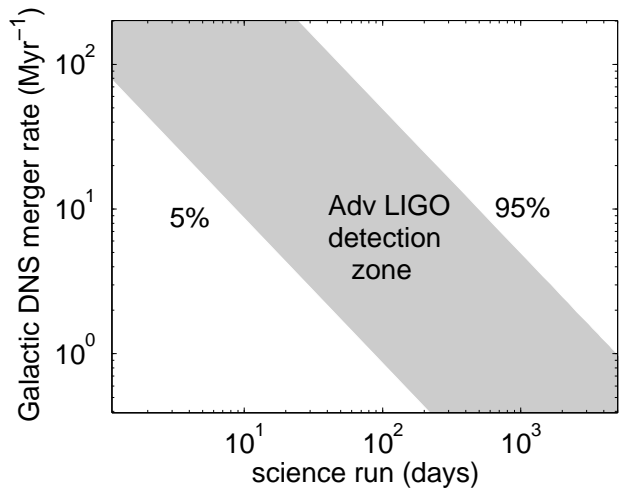}
\caption{Plot of the detection zone, assuming an Advanced LIGO horizon distance of 450 Mpc, with Galactic DNS merger rates (vertical axis) spanning $1-100$ Myr$^{-1}$ per Milky Way galaxy. The intersection of a particular rate with the 5\% and 95\% probability contours gives the waiting times. The corresponding waiting times for the upper and lower rates at the 5\% level are 0.5 and 90 days respectively.  }
\end{figure}\label{fig4}

\section{Results and Discussion}

Figure 2 plots the detection zone defined by equation (\ref{ezone}). The $D_{5\%}^{\mathrm{PEH}}$ curve shows that the inhomogeneity of the local Universe affects the PEH at horizon distances less than 50 Mpc. Waiting times corresponding to $D_{\mathrm{h}}<50$ Mpc are reduced compared with a uniform galaxy distribution (dot-dashed line) but the deviation manifests after relatively long observation times i.e. $5-10$ yr. For a horizon distance corresponding to a sensitivity of Enhanced LIGO, $\sim 60$ Mpc, the waiting time is about 1 yr and the detection zone spans about 60 yr. Assuming a horizon distance of Advanced LIGO, the waiting time is about 1 day and the detection zone spans 60 days. This implies that there is a 90\% probability of observing at least one DNS merger in $1-60$ days of a science run at Advanced LIGO sensitivities. 

Figure 3 plots the fractional difference between the PEH$_{5\%}$ contours for a model that assumes a uniform galaxy distribution and one that includes inhomogeneities in galaxy density in the local Universe. There is a maximum factor of about 1.4 decrease in the horizon distance from clustering in the local Universe, corresponding to an observation time of 7--10 years. Considering that Enhanced and Advanced LIGO should have horizon distances out to the nearly homogenous galaxy space density regime, it is clear that this effect will not be important for science runs of practical duration.

\cite{kopp08} show that the relatively small horizon distances of the two initial LIGO detectors can vary because of the relative orientation of the detector to the sky positions of galaxies in the local Universe. This is because the detector antenna pattern must coincide with the sky position of the DNS host galaxies for a detection. The galaxies become sparser and the sky separations become larger as the horizon distance becomes smaller. In this analysis, the effect of different detector locations has not been included, as this study focusses more on detector sensitivities comparable to those of Enhanced and Advanced LIGO. The associated horizon distances of these detectors extend from the border of inhomogenous to mostly uniform galaxy space densities.    

Figure 4 plots the detection zone at a fixed horizon distance corresponding to that of Advanced LIGO, but with a varying mean DNS merger rate of $1-100$ Myr$^{-1}$ per Milky Way Galaxy to account for the uncertainty in this rate. The corresponding minimum waiting times at the 5\% probability level for the two rate extremes are 0.5 and 90 days. 

In summary, inclusion of the Poisson uncertainty for the minimum waiting times of DNS mergers detectable by GW observatories provides a more realistic means for defining a detection zone in terms of the duration of a science run. Even if the rate density of DNS mergers is well constrained, the study demonstrates how the waiting times can vary widely over an observation period. To highlight this, the model calculates a 1/20 chance of observing a DNS merger by Enhanced LIGO in only 1 yr of observation assuming an event rate of about 80 Myr$^{-1}$ in the Milky Way Galaxy. The horizon distance of Enhanced LIGO is also in the distance regime where inhomogeneities in galaxy space density will only marginally influence the probability of a detection. 

The waiting times and zone width at an Advanced LIGO sensitivity are much shorter in duration and imply a first detection at the 95\% probability level will occur in less than 60 days and there is a 1/20 chance of a first detection in about 1 day. If the merger rate is of order 1 Myr$^{-1}$ in the Milky Way Galaxy, the waiting times at a 5\% probability level assuming an Advanced LIGO horizon distance are extended to about 90 days.

\section*{Acknowledgments}
D. M. Coward is supported by an Australian Research Council grant (LP0667494). This research was made possible by a visit to the Observatoire de la C\^{o}te d'Azur, funded by the Australian Academy of Science: Scientific visits to Europe program. The host institute provided a fertile environment for discussions with Dr Tania Regimbau that stimulated this work. The author also thanks Dr Ron Burman, Prof. David Blair for fruitful discussions and Prof. Richard O'Shaughnessy for providing a review on behalf of the LIGO Scientific Community. Finally, the author thanks the anonymous  referee for providing very informative comments and suggestions for improving this paper.

\label{lastpage}

\end{document}